\documentclass[12pt]{article}
\usepackage{amssymb,amsmath,epsfig,cite}
\allowdisplaybreaks

\begin{document}
\title{\bf Electromagnetic Effects on the Complexity of Static Cylindrical Object in $f(G,T)$
Gravity}
\author{M. Sharif$^1$ \thanks{msharif.math@pu.edu.pk} and K. Hassan$^2$ \thanks{komalhassan3@gmail.com}\\
$^1$ Department of Mathematics and Statistics, The University of Lahore,\\
1-KM Defence Road Lahore, Pakistan.\\
$^2$ Department of Mathematics, University of the Punjab,\\
Quaid-e-Azam Campus, Lahore-54590, Pakistan.}

\date{}
\maketitle

\begin{abstract}
In this paper, we investigate complexity of anisotropic cylindrical
object under the influence of electromagnetic field in $f(G,T)$
theory, where $G$ and $T$ indicate the Gauss-Bonnet term and trace
of the stress-energy tensor, respectively. For this purpose, we
calculate the modified field equations, non-conservation equation
and mass distributions that assist in comprehending the structure of
astrophysical objects. The Riemann tensor is divided into different
structure scalars, among which one is called the complexity factor.
This factor is used to measure complexity of the system due to the
involvement of inhomogeneous energy density, anisotropic pressure
and charge. The vanishing of the complexity factor is employed as a
constraint to formulate charged static solutions for the
Gokhroo-Mehra model and polytropic equation of state. We conclude
that the presence of charge reduces the complexity of the
anisotropic system.
\end{abstract}

\section{Introduction}

General theory of relativity (GR) improved the fundamental ideas
about our universe and the notion of gravity. Einstein assumed that
the universe was neither contracting nor expanding, so he considered
it as a static epoch. Later, Hubble confirmed the expansion of the
cosmos by relating the recession velocity of galaxies and distance.
The inexplicable force known as dark energy is thought to be
responsible for the accelerated expansion of the universe.
Supernovae and microwave background radiations are among different
astrophysical events that unveil the fast cosmic expansion
\cite{1a}. The $\Lambda$CDM model is the best model to explain the
current accelerated expansion of the universe, however, it has two
major challenges: fine-tuning and cosmic coincidence. Thus, some
modifications in GR (so called modified theories of gravity) were
introduced to explain the expanding universe.

In this regard, Nojiri and Odintsov \cite{4} revised the
Einstein-Hilbert action by adding the function of Gauss-Bonnet (GB)
invariant, which resulted in $f(G)$ gravity. This also helped in
explaining how the cosmos transitioned from a decelerating to an
accelerating epoch. The GB invariant is a combination of the Riemann
tensor $(R_{\psi\chi\beta\alpha})$, Ricci tensor $(R_{\psi\chi})$
and curvature scalar $(R)$ as
\begin{equation*}
G=-4R^{\psi\chi}R_{\psi\chi}+R^{\psi\chi\beta\alpha}R_{\psi\chi\beta\alpha}+R^2.
\end{equation*}
Bamba et al. \cite{5} studied the early-time bounce as well as
late-time acceleration by considering the scale factor in the form
of an exponential function by reconstructing $f(G)$ model. Abbas et
al. \cite{5a} examined viable and stable aspects of stellar objects
by utilizing the Krori-Barua metric together with a power-law model
in $f(G)$ theory. Sharif and Ikram \cite{5b} addressed the
inflationary characteristics of homogeneous and isotropic universe
using feasible $f(G)$ model. For a plane symmetric structure, Shamir
and Saeed \cite{6} investigated the exponential and power-law
solutions in this theory. Sharif and Ramzan \cite{6b} applied the
embedding class-1 method to examine physical features of anisotropic
self-gravitating objects.

Sharif and Ikram \cite{7} introduced trace of the energy-momentum
tensor in $f(G)$ action and proposed a new modified theory named as
$f(G,T)$ theory. They discussed energy conditions using FRW metric
in this theory. The same authors \cite{7a} used the perturbation
approach to check the stable behavior of the Einstein universe. For
a certain range of the parameters, Hossienkhani et al. \cite{7b}
demonstrated that weak and null energy conditions show consistency
with anisotropic $f(G,T)$ universe. Sharif and Naeem \cite{8}
developed a new solution to analyze viability and stability of
different stellar configurations. Shamir \cite{8a} investigated the
oscillating universe for two models, such as linear and logarithmic
trace terms of $f(G,T)$ gravity.

In the evolution and stable configuration of the celestial entities,
electromagnetic field plays a crucial role as it suppresses the
gravitational force. A significant quantity of charge is required to
counteract gravitational attraction and maintain the stabilizing
behavior of self-gravitating entities. In astrophysical realm, the
first static charged solution of the Einstein-Maxwell field
equations was presented by Reissner and Nordstrom. Bondi \cite{9}
imposed the known cylindrical wave solution of the free space on the
static field and observed the propagation of energy from such waves.
Ivanov \cite{9a} studied general and regular solutions of the field
equations in three different classes for the charged cylindrical
object. Esculpi and Aloma \cite{9ab} investigated the influence of
anisotropy on charged compact objects using linear equation of state
that relates radial and tangential pressures. Sharif and Naz
\cite{9b} used dynamical field equations to investigate the
collapsing phenomenon of a cylindrically charged star and evaluated
the relationship between GB terms, physical variables and Weyl
tensor in $f(G)$ gravity. We have worked on Tolman IV and
Krori-Barua spacetimes to illustrate physical features of the
compact object in uncharged/charged configurations, respectively, in
$f(G,T)$ theory \cite{9c}.

The complexity of self-gravitating objects is greatly influenced by
feasible characteristics of astronomical entities, such as
temperature, heat, pressure, energy density, etc. Therefore, a
mathematical formula that includes all significant components must
be used to determine complexity of cosmic entities. The notion of
complexity in terms of entropy and information was introduced by
L{\'o}pez-Ruiz et al. \cite{10a}. Initially, this concept was
implemented on two physical objects, i.e., perfect crystal and ideal
gas. The molecules of perfect crystal are grouped in an ordered
manner (due to the symmetric nature), thus it has zero entropy. In
contrast, as the molecules of ideal gas are scattered randomly, so
it shows maximum entropy. The probability distribution around the
symmetric component of a perfect crystal adds little information
because of its symmetry, and a small portion is enough to
characterize all of its properties. However, maximum amount of
information is available while studying a tiny region of ideal gas.
Both these structures exhibit different behavior, yet they are
allocated zero complexity. Another appropriate concept of complexity
was established on the basis of how different probabilistic states
depart from the equiprobable distribution of the structure
\cite{10b}. Using this concept, both ideal gas and perfect crystal
are considered to give zero complexity. The energy density was
employed in the revised methodology (in place of probability
distribution) to evaluate the complexity of celestial objects
\cite{11a}. However, other state variables like heat, pressure,
temperature, etc. were not included in this benchmark.

Recently, Herrera \cite{12} redefined the notion of complexity in
terms of inhomogeneous energy density, pressure anisotropy and
Tolman mass for the case of static anisotropic fluid source. In this
approach, different structure scalars are obtained through the
orthogonal breakdown of the Riemann tensor. A scalar that
encompasses all the above-mentioned parameters is called complexity
factor. Sharif and Butt \cite{16} studied how the complexity of
spherical configuration was affected in the presence of an
electromagnetic field. Herrera and his collaborators \cite{13}
extended the concept of complexity for dynamical dissipative
configuration and addressed two modes of evolution. The same authors
\cite{14} also considered an axially symmetric structure to
calculate its complexity. Contreras et al. \cite{15a} utilized the
temporal metric component of Durgapal IV and V solutions along with
vanishing complexity condition for both charged/uncharged cases to
discuss feasibility of the constructed models. Contreras and
Stuchlik \cite{15b} formulated a simple protocol to develop the
anisotropic interior solutions via vanishing condition. Abbas and
Nazar \cite{15c} calculated the complexity of various structures in
the realm of non-minimally coupled $f(R)$ gravity. Sharif et al.
\cite{17a} applied Herrera's approach to the spherically symmetric
matter source in $f(G)$ gravity and found an increment in
complexity. A large body of literature is found on the application
of complexity to different geometries in the background of various
modified theories \cite{18}. Yousaf et al. \cite{17b} inspected the
complexity producing factor for the charged/uncharged spheres in
$f(G,T)$ gravity.

When a strong gravitational field is present, the departure from
spherical to non-spherical geometries becomes crucial for assessing
the solution. Levi Civita developed vacuum cylindrically symmetric
spacetime that inspired astronomers to explore the interesting
features of various stellar systems. In order to examine the
cylindrical distribution, Herrera et al.\cite{17d} discussed scalar
functions in the form of matter variables. Using modified
Gauss-Bonnet gravity, Houndjo et al. \cite{17f} constructed a set of
seven solutions that conform to three different viable models in
cylindrical spacetime. To determine the complexity of a cylindrical
spacetime, Sharif and Butt \cite{15} applied Herrera method to
decompose the Riemann tensor in the scenario of static matter
configuration. The same authors \cite{17} examined the influence of
electromagnetic field on the cylindrical geometry using the same
procedure. We have also investigated the complexity factor in static
cylindrical and dynamical spherical/cylindrical structures in
$f(G,T)$ gravity \cite{17g}.

This paper studies the effects of charge on the complexity of
anisotropic cylindrical matter distribution in the formalism of
$f(G,T)$ theory. Throughout the paper, we have utilized the
geometrized units, i.e., $G=c=1$. The paper is characterized as
follows. The modified field equations and physical characteristics
related to anisotropic matter distribution are derived in section
\textbf{2}. The orthogonal splitting of the Riemann tensor is given
in section \textbf{3}. In section \textbf{4}, the vanishing
complexity condition is employed as a constraint to compute the
possible solutions. The key results are wrapped in section
\textbf{5}.

\section{$f(G,T)$ Gravity}

In this section, we explore the modified field equations with the
inclusion of electromagnetic field and examine physical attributes
of the anisotropic celestial structure. The Einstein-Maxwell action
is given as follows
\begin{equation}\label{1}
\mathrm{I}_{f(G,T)}=\frac{1}{2\kappa^2}\int\sqrt{-g}[f(G,T)+R]d^{4}x+\int\sqrt{-g}(\mathfrak{L}_{m}+\mathfrak{L}_{\textsf{EM}})d^{4}x,
\end{equation}
where $g$ denotes determinant of the metric tensor and $\kappa^{2}$
represents the coupling constant. Moreover, the terms
$\mathfrak{L}_{m}$ and $\mathfrak{L}_{\textsf{EM}}$ indicate the
Lagrangian density corresponding to the usual matter and
electromagnetic field, respectively. The energy-momentum tensor and
matter Lagrangian density are associated through the relation
\begin{equation}\label{2}
T_{\psi\chi}=g_{\psi\chi}\mathfrak{L}_{m}-\frac{2\partial\mathfrak{L}_{m}}{\partial
g^{\psi\chi}}.
\end{equation}
The $f(G,T)$ field equations are obtained by varying the action
\eqref{1} with respect to the metric tensor
\begin{eqnarray}\nonumber
G^{\psi}_{\chi}&=&8\pi
(T^{\psi}_{\chi}+\textsf{S}^{\psi}_{\chi})-(\Theta^{\psi}_{\chi}+T^{\psi}_{\chi})f_{T}(G,T)+\frac{1}{2}\delta^{\psi}_{\chi}f(G,T)+\left(4R_{\chi
m}R^{\psi m}\right.\\\nonumber &-&\left.2RR^{\psi}_{\chi}+4R^{m
l}R^{\psi}_{m \chi l}-2R^{m l n \psi}R_{\chi m l
n}\right)f_{G}(G,T)+\left(4R^{\psi}_{\chi}\nabla^{2}\right.\\\nonumber
&-&\left.4R^{\psi}_{m \chi
l}\nabla^{m}\nabla^{l}-2\delta^{\psi}_{\chi}R\nabla^{2}+2R\nabla^{\psi}\nabla_{\chi}
-4R^{\psi
m}\nabla_{m}\nabla_{\chi}-4R^{m}_{\chi}\nabla^{\psi}\nabla_{m}\right.\\\label{3}
&+&\left.4\delta^{\psi}_{\chi}R^{m
l}\nabla_{m}\nabla_{l}\right)f_{G}(G,T),
\end{eqnarray}
where the Einstein tensor is expressed as
$G^{\psi}_{\chi}=R^{\psi}_{\chi}-\frac{1}{2}R\delta^{\psi}_{\chi}$
and $\nabla^{\upsilon}\nabla_{\upsilon}=\nabla^{2}$ denotes the d'
Alembert operator. Moreover, the partial derivatives of an arbitrary
$f(G,T)$ function such as $f_{G}(G,T)$ and $f_{T}(G,T)$ represent
derivatives taken with respect to $G$ and $T$, respectively, and
$\Theta_{\psi\chi}=-2T_{\psi\chi}+pg_{\psi\chi}$.

The cylindrical geometric distribution surrounded by the
hypersurface $(\Sigma)$ is delineated by the line element \cite{15}
\begin{equation}\label{7}
ds^{2}=-K^2{(r)}dt^{2}+L^2(r)dr^{2}+r^2d\theta^{2}+r^2\alpha^2{dz^2},
\end{equation}
where the metric coefficients $K$ and $L$ are unknown potential
functions while $\alpha$ (constant term) has the dimension of
inverse length. The stress-energy tensor corresponding to
anisotropic matter distribution reads
\begin{equation}\label{5}
T^{\psi M}_{\chi}=\varrho
\mathrm{u}^{\psi}\mathrm{u}_{\chi}+ph^{\psi}_{\chi}+\Pi^{\psi}_{\chi},
\end{equation}
along with
\begin{eqnarray*}\nonumber
\Pi^{\psi}_{\chi}&=&\Pi\left(\phi^{\psi}\phi_{\chi}-\frac{1}{3}h^{\psi}_{\chi}\right),
\quad \Pi=p_{r}-p_{\bot},\quad p=\frac{p_{r}+2p_{\bot}}{3},\\
\mathrm{u}^{\psi}\mathrm{u}_{\psi}&=&-1, \quad
\phi^{\psi}\mathrm{u}_{\psi}=0,\quad \phi^{\psi}\phi_{\psi}=1\quad
h^{\psi}_{\chi}=\delta^{\psi}_{\chi}+\mathrm{u}^{\psi}\mathrm{u}_{\chi},
\end{eqnarray*}
where $p_{\bot}$ and $p_{r}$ are the tangential and radial
pressures, respectively, and $\Pi$ is the anisotropic factor. The
components of four velocity and radial four-vector are denoted as
$\mathrm{u}^{\psi}=\left(\frac{1}{K},0,0,0\right)$,
$\phi^{\psi}=\left(0,\frac{1}{L},0,0\right)$, respectively. The
extra curvature terms of $f(G,T)$ gravity are given as
\begin{eqnarray}\nonumber
T^{\psi
GT}_{\chi}&=&\frac{1}{8\pi}\left[\{(\varrho+p)\mathrm{u}^{\psi}\mathrm{u}_{\chi}+\Pi^{\psi}_{\chi}\}f_{T}(G,T)+\frac{1}{2}\delta^{\psi}_{\chi}f(G,T)+\left(4R_{\chi
m}R^{\psi m}\right.\right.\\\nonumber &+&\left.\left.4R^{m
l}R^{\psi}_{m \chi l}-2RR^{\psi}_{\chi}-2R^{m l n \psi}R_{\chi m l
n}\right)f_{G}(G,T)+\left(4\delta^{\psi}_{\chi}R^{m
l}\nabla_{m}\nabla_{l}\right.\right.\\\nonumber
&+&\left.\left.4R^{\psi}_{\chi}\nabla^{2}+2R\nabla^{\psi}\nabla_{\chi}-2\delta^{\psi}_{\chi}R\nabla^{2}
-4R^{\psi m}\nabla_{m}\nabla_{\chi}-4R^{\psi}_{m \chi
l}\nabla^{m}\nabla^{l}\right.\right.\\\label{6}
&-&\left.\left.4R^{m}_{\chi}\nabla^{\psi}\nabla_{m}\right)f_{G}(G,T)\right].
\end{eqnarray}
In order to investigate charged matter distribution, the
electromagnetic energy tensor is defined as
\begin{equation}\label{3b}
\textsf{S}_{\psi\chi}=\frac{1}{4\pi}\left(F^{l}_{\psi}F_{\chi
l}-\frac{1}{4}F_{lm}F^{lm}g_{\psi\chi}\right),
\end{equation}
where $F_{\psi\chi}=\gamma_{\chi,\psi}-\gamma_{\psi,\chi}$ stands
for the Maxwell field tensor and $\gamma_{\psi}$ is the four
potential which becomes $\gamma_{\psi}=\gamma(r)\delta^{0}_{\psi}$
for the static structure. The Maxwell field equations, in the
tensorial notation, read
\begin{equation}\nonumber
F^{\psi\chi}_{~~;\chi}=4\pi \textsf{J}^{\psi} ,\quad
F_{[\psi\chi;l]}=0,
\end{equation}
where $\textsf{J}^{\psi}=\varpi \upsilon^{\psi}$ indicates the
electromagnetic four-current vector and $\varpi$ represents the
charge density.

In this framework, the modified field equations are analyzed as
\begin{eqnarray}\label{8}
8\pi\big(\varrho^{\textsf{eff}}+\frac{\textsf{q}^2}{8\pi
r^4}\big)&=&\left[\frac{1}{rL^2}\left(\frac{2L'}{L}-\frac{1}{r}\right)\right],
\\\label{9}
8\pi \big(p_{r}^{\textsf{eff}}-\frac{\textsf{q}^2}{8\pi r^4}\big)
&=&
\left[\frac{1}{rL^2}\left(\frac{2K'}{K}+\frac{1}{r}\right)\right],\\\label{10}
8\pi \big(p_{\bot}^{\textsf{eff}}+\frac{\textsf{q}^2}{8\pi
r^4}\big)&=&\frac{K'}{rKL^2}+\frac{K''}{KL^2}-\frac{K'L'}{KL^3}-\frac{L'}{rL^3},
\end{eqnarray}
where derivative with respect to $r$ is signified by prime and the
correction terms $\varrho^{\textsf{eff}}$, $p_{r}^{\textsf{eff}}$
and $p_{\bot}^{\textsf{eff}}$ are computed as
\begin{eqnarray*}
\varrho^{\textsf{eff}}&=&\varrho+\frac{1}{8\pi}\left[(\varrho+p)f_{T}-\frac{f}{2}+\frac{4K''f_{G}}{r^2KL^4}+\frac{12L'f_{G}'}{r^2L^5}
-\frac{12K'L'f_{G}}{r^2KL^5}-\frac{4f_{G}''}{r^2L^4}\right],\\\nonumber
p_{r}^{\textsf{eff}}&=&p_{r}+\frac{1}{8\pi}\left[\frac{2}{3}\Pi
f_{T}+\frac{f}{2}+\frac{12K'L'f_{G}}{r^2KL^5}+\frac{12K'f_{G}'}{r^2KL^4}-\frac{4K''f_{G}}{r^2KL^4}
\right],\\\nonumber
p_{\bot}^{\textsf{eff}}&=&p_{\bot}+\frac{1}{8\pi}\left[-\frac{\Pi}{3}
f_{T}+\frac{f}{2}-\frac{4K''f_{G}}{r^2KL^4}-\frac{12K'L'f_{G}'}{rKL^5}+\frac{4K'f_{G}''}{rKL^4}
+\frac{12K'L'f_{G}}{r^2KL^5}\right.\\\nonumber&+&\left.\frac{4K''f_{G}'}{rKL^4}
\right].
\end{eqnarray*}
The non-conservation of energy-momentum tensor is observed through
the covariant differentiation of Eq.\eqref{3}. As a result, an extra
force appears that pushes the particles to disobey the geodesic
trajectory. The corresponding non-conservation equation as well as
the hydrostatic equilibrium equation, respectively, under the effect
of electromagnetic field are written as
\begin{eqnarray}\nonumber
\nabla^{\psi}T_{\psi\chi}&=&\frac{f_{T}(G,T)}{k^{2}-f_{T}(G,T)}\left[(T_{\psi\chi}+\Theta_{\psi\chi})\nabla^{\psi}(\ln
f_{T}(G,T))+\nabla^{\psi}\Theta_{\psi\chi}
\right.\\\label{11a}&-&\left.\frac{1}{2}g_{\psi\chi}\nabla^{\psi}T\right],\\\label{11}
(p^{eff}_{r}-\frac{\textsf{q}^2}{8\pi r^4})^{'}&=&-\frac{K'
(p^{\textsf{eff}}_{r}+\varrho^{\textsf{eff}})}{K}+\frac{2(p^{\textsf{eff}}_{\bot}-p^{\textsf{eff}}_{r}+\frac{\textsf{q}^2}{4\pi r^4})}{r}+
\textsf{Z}L^2,
\end{eqnarray}
where the curvature terms present in \textsf{Z} are displayed in
Appendix \textbf{A}. The Tolman-Oppenheimer-Volkoff equation can
also be represented by Eq.\eqref{11} that helps to describe the
formation of anisotropic geometry. Equation \eqref{9} provides the
value of $\frac{K'}{K}$ as
\begin{equation}\label{12}
\frac{K'}{K}=\frac{4r^2}{r^2-8mr+4\textsf{q}^2}\left(4\pi
r(p^{\textsf{eff}}_r-\frac{\textsf{q}^2}{8\pi r^4})-\frac{1}{8r}-\frac{\textsf{q}^2}{2r^3}+\frac{m}{r^2}\right).
\end{equation}
Utilizing the above expression in Eq.(11), we have
\begin{align}\nonumber
\bigg(p^{eff}_{r}-\frac{\textsf{q}^2}{8\pi
r^4}\bigg)^{'}&=-\frac{4r^2}{r^2-8mr+4\textsf{q}^2}\left(4\pi
r(p^{\textsf{eff}}_r-\frac{\textsf{q}^2}{8\pi
r^4})-\frac{1}{8r}-\frac{\textsf{q}^2}{2r^3}+\frac{m}{r^2}\right)\\\label{13}&\times\left(\varrho^{\textsf{eff}}+p^{\textsf{eff}}_r\right)+\frac{2(p^{\textsf{eff}}_{\bot}-p^{\textsf{eff}}_{r}+\frac{\textsf{q}^2}{4\pi
r^4})}{r}+ \textsf{Z}L^2,
\end{align}
where $m$ is the mass of the fluid distribution. The mass function
of the interior geometry can be evaluated through C-energy and
Tolman mass. Firstly, the C-energy formalism \cite{19b} is utilized
to compute the mass as
\begin{equation}\label{14}
m(r)\cong
\widetilde{\texttt{E}}=l{\texttt{E}}=\left(\frac{1}{4}-\frac{1}{L^2}\right)\frac{r}{2}+\frac{\textsf{q}^2}{2r},
\end{equation}
Making use of the field Eq.\eqref{8}, it can be rewritten as
\begin{equation}\label{15}
m=\int_{0}^{r}4\pi{r}^{2}\bigg(\varrho^{\textsf{eff}}+\frac{\textsf{q}^2}{8\pi
r^4}\bigg)d{r}+\frac{\textsf{q}^2}{2r}+\frac{r}{8}.
\end{equation}

The Weyl tensor calculates the fluctuation in the celestial system
caused by the gravitational fields in nearby objects. The Riemann
tensor is divided into the Ricci tensor, curvature scalar and Weyl
tensor as
\begin{equation}\label{18}
R^{\sigma}_{\psi\chi \omega}=C^{\sigma}_{\psi\chi
\omega}+\frac{1}{2}R_{\psi \omega}\delta^{\sigma}_{\chi}
-\frac{1}{2}R_{\psi\chi}\delta^{\sigma}_{\omega}+\frac{1}{2}R^{\sigma}_{\chi}g_{\psi
\omega}-\frac{1}{2}R^{\sigma}_{\omega}g_{\psi\chi}
-\frac{1}{6}R\bigg(\delta^{\sigma}_{\chi}g_{\psi
\omega}-g_{\psi\chi}\delta^{\sigma}_{\omega}\bigg).
\end{equation}
The decomposition of the Weyl tensor through the observer's four
velocity gives magnetic and electric parts as
\begin{equation}\nonumber
H_{\psi\chi}=\frac{1}{2}\eta_{\psi\lambda\beta\gamma}C^{\beta\gamma}_{\chi
\rho}\mathrm{u}^{\lambda}\mathrm{u}^{\rho} ,\quad
\textsf{E}_{\psi\chi}=C_{\psi\gamma\chi\delta}\mathrm{u}^{\gamma}\mathrm{u}^{\delta},
\end{equation}
where
\begin{equation}\label{19}
C_{\psi\chi\lambda\kappa}=\left(g_{\psi\chi\beta\alpha}g_{\lambda\kappa\gamma\delta}-\eta_{\psi\chi\beta\alpha}\eta_{\lambda\kappa\gamma\delta}\right)
\mathrm{u}^{\beta}\mathrm{u}^{\gamma}\textsf{E}^{\alpha\delta},
\end{equation}
and $\eta_{\psi\chi\beta\alpha}$ indicates the Levi-Civita tensor
along with
$g_{\psi\chi\beta\alpha}=g_{\psi\beta}g_{\chi\alpha}-g_{\psi\alpha}g_{\chi\beta}$.
The magnetic part of the Weyl tensor vanishes in symmetric
cylindrical spacetime, while the electric component becomes
\begin{equation}\nonumber
\textsf{E}_{\psi\chi}=\textsf{E}\left(\phi_{\psi}\phi_{\chi}-\frac{h_{\psi\chi}}{3}\right),
\end{equation}
where $\textsf{E}$ and some of its properties are
\begin{equation}\label{20}
\textsf{E}=\frac{1}{2KL^2}\bigg[K''-\frac{K'L'}{L}+\frac{KL'}{rL}-\frac{K'}{r}+\frac{K}{r^2}\bigg],
\quad \textsf{E}^{\psi}_{\psi}=0=\textsf{E}_{\psi\gamma}\mathrm{u}^{\gamma}=0.
\end{equation}

The connection between the mass of the cylindrical object with
matter variables, charge, and Weyl tensor is given by
Eqs.\eqref{8}-\eqref{10}, \eqref{15} and \eqref{20} as
\begin{equation}\label{21}
m(r)=\frac{4\pi
r^{3}}{3}\left(\varrho^{\textsf{eff}}+p^{\textsf{eff}}_{\bot}-p^{\textsf{eff}}_{r}+\frac{3\textsf{q}^2}{8\pi
r^4}\right)+\frac{\textsf{q}^2}{2r}-\frac{\textsf{E}r^{3}}{3}+\frac{r}{8},
\end{equation}
and the value of $\textsf{E}$ from the above equation leads to
\begin{equation}\label{22}
\textsf{E}=\frac{4\pi}{r^{3}}\int_{0}^{r}r^{3}\bigg(\varrho^{\textsf{eff}}+\frac{\textsf{q}^2}{8\pi
r^4}\bigg)^{'}d{r}-4\pi\left(p^{\textsf{eff}}_{r}-p^{\textsf{eff}}_{\bot}-\frac{\textsf{q}^2}{4\pi
r^4}\right).
\end{equation}
This shows the relation among the energy density inhomogeneity,
pressure anisotropy, electromagnetic effect and the Weyl tensor in
the presence of $f(G,T)$ theory. After inserting this value of
$\textsf{E}$ in Eq.\eqref{21}, we have
\begin{equation}\label{23}
m(r)=\frac{\textsf{q}^2}{2r}+\frac{4\pi
r^{3}}{3}\bigg(\varrho^{\textsf{eff}}+\frac{\textsf{q}^2}{8\pi
r^4}\bigg)-\frac{4\pi}{3}\int_{0}^{r}{r}^{3}\bigg(\varrho^{\textsf{eff}}+\frac{\textsf{q}^2}{8\pi
r^4}\bigg)^{'}d{r}+\frac{r}{8}.
\end{equation}
The effects of inhomogeneous energy density and charge on the mass
can be noticed from the above equation.

For the static symmetric structure, the total mass of the fluid
distribution is determined by the Tolman formula \cite{22a} as
\begin{equation}\label{24}
m_{T}=4\pi\int^{r}_{0}KLr^{2}\left(-T^{0\textsf{eff}}_{0}+T^{1\textsf{eff}}_{1}+2T^{2\textsf{eff}}_{2}\right)dr.
\end{equation}
Using the field equations in the above equation, it follows that
\begin{equation}\label{25}
m_{T}=\frac{K'}{L}r^2.
\end{equation}
Inserting the value of $K^{'}$ from Eq.\eqref{12}, it follows that
\begin{equation}\nonumber
m_T=KL\big(m-\frac{\textsf{q}^2}{2r}\big)+4\pi
KL\big(p^{\textsf{eff}}_{r}-\frac{\textsf{q}^2}{8\pi
r^4}\big)r^3-\frac{KLr}{8}.
\end{equation}
The Tolman mass can also be referred to as the effective
gravitational mass as it interlinks with the gravitational
acceleration of test particles by
\begin{equation}\label{26}
a=\frac{K'}{KL}=\frac{m_T}{Kr^2}.
\end{equation}
We obtain the expression for $m_{T}$ \cite{23a} as
\begin{align}\nonumber
m_{T}&=(m_{T})_{\Sigma}\left(\frac{r}{r_{\Sigma}}\right)^{3}-r^{3}\int^{r_{\Sigma}}_{r}\frac{KL}{r^4}\bigg[8\pi
r^3\left(p^{\textsf{eff}}_{\bot}+\frac{\textsf{q}^2}{8\pi
r^4}-p^{\textsf{eff}}_{r}+\frac{\textsf{q}^2}{8\pi
r^4}\right)\\\label{27}&+\int^{r}_{0}4\pi
{r}^{3}(\varrho^{\textsf{eff}}+\frac{\textsf{q}^2}{8\pi
r^4})^{'}dr\bigg]d{r}.
\end{align}
Alternatively, the Tolman mass is acquired in terms of $\textsf{E}$
as
\begin{equation}\label{28}
m_{T}=\left(\frac{r}{r_{\Sigma}}\right)^{3}(m_{T})_{\Sigma}-r^{3}\int^{r_{\Sigma}}_{r}\frac
{KL}{r^4}\left[4\pi
r^3\left(p^{\textsf{eff}}_{\bot}+\frac{\textsf{q}^2}{8\pi
r^4}-p^{\textsf{eff}}_{r}+\frac{\textsf{q}^2}{8\pi
r^4}\right)+r^3\textsf{E}\right]d{r}.
\end{equation}
It is important to notice that inhomogeneous energy density,
electromagnetic field, correction terms, and anisotropic pressure
affect the Tolman mass.

\section{Structure Scalars}

Bel \cite{22b} was the pioneer to propose the orthogonal splitting
of the Riemann tensor, followed by Herrera \cite{24a}, who
formulated various scalar functions known as structure scalars.
These scalar functions are defined on account of the essential
characteristics of fluid distributions such as inhomogeneous energy
density, anisotropic pressure, electromagnetic field and active
gravitational mass. These structure scalars are crucial in
comprehending the complexity factor for the self-gravitating
systems. The tensors $\textsf{Y}_{\psi\chi}$,
$\textsf{Z}_{\psi\chi}$ and $\textsf{X}_{\psi\chi}$ are defined as
\begin{eqnarray}\label{29}
\textsf{Y}_{\psi\chi} &=&R_{\psi\gamma\chi\delta}\mathrm{u}^{\gamma}\mathrm{u}^{\delta},  \\
\textsf{Z}_{\psi\chi}
&=&_{\ast}R_{\psi\gamma\chi\delta}\mathrm{u}^{\gamma}\mathrm{u}^{\delta}=
\frac{1}{2}\eta_{\psi\gamma\epsilon\beta}R^{\epsilon\beta}_{\chi\delta}\mathrm{u}^{\gamma}\mathrm{u}^{\delta},  \\
\textsf{X}_{\psi\chi}
&=&^{\ast}R^{\ast}_{\psi\gamma\chi\delta}\mathrm{u}^{\gamma}\mathrm{u}^{\delta}=\frac{1}{2}\eta^{\epsilon\beta}_{\psi\gamma},
R^{\ast}_{\epsilon\beta\chi\delta}\mathrm{u}^{\gamma}\mathrm{u}^{\delta},
\end{eqnarray}
where
$R^{\ast}_{\psi\chi\gamma\delta}=\frac{1}{2}\eta_{\epsilon\alpha\gamma\delta}R^{\epsilon\alpha}_{\psi\chi}$.
Equation (\ref{18}) can also be rewritten in terms of the Riemann
tensor as
\begin{equation}\label{30}
R^{\psi\gamma}_{\chi\delta}=C^{\psi\gamma}_{\chi\delta}+16\pi
T^{\textsf{eff}}{[\psi}_{[\chi}\delta^{\gamma]}_{\delta]}+8\pi
T^{\textsf{eff}}\left(\frac{1}{3}\delta^{\psi}_{[\chi}\delta^{\gamma}_{\delta]}-\delta^{[\psi}_{[\chi}\delta^{\gamma]}_{\delta]}\right).
\end{equation}
The above equation is decomposed into four tensorial quantities
after substituting Eqs.(\ref{5}) and (\ref{6}) as
\begin{eqnarray}\nonumber
R^{\psi\gamma}_{(I)\chi\delta} &=&
16\pi\varrho\mathrm{u}^{[\psi}\mathrm{u}_{[\chi}\delta^{\gamma]}_{\delta]}+2\varrho\mathrm{u}^{[\psi}\mathrm{u}_{[\chi}\delta^{\gamma]}_{\delta]}
+16\pi
ph^{[\psi}_{[\chi}\delta^{\gamma]}_{\delta]}\\\label{70a}&+&2p
\mathrm{u}^{[\psi}
\mathrm{u}_{[\chi}\delta^{\gamma]}_{\delta]}+8\pi(\varrho+3p)\left(\frac{1}{3}\delta^{\psi}_{[\chi}\delta^{\gamma}_{\delta]}-
\delta^{[\psi}_{[\chi}\delta^{\gamma]}_{\delta]}\right),
\\\nonumber
R^{\psi\gamma}_{(II)\chi\delta} &=&
16\pi\Pi^{[\psi}_{[\chi}\delta^{\gamma]}_{\delta]}+2\Pi^{[\psi}_{[\chi}\delta^{\gamma]}_{\delta]}+
\delta^{[\psi}_{[\chi}\delta^{\gamma]}_{\delta]}f\\\label{71a}&+&8\delta^{[\psi}_{[\chi}\delta^{\gamma]}_{\delta]}
R^{ml}\nabla_{m}\nabla_{l}f_{G}-4R\delta^{[\psi}_{[\chi}\delta^{\gamma]}_{\delta]}\Box
f_{G},  \\\label{72a} R^{\psi\gamma}_{(III)\chi\delta} &=&
4\varrho\chi^{[\psi}_{[\chi}\textsf{E}^{\gamma]}_{\delta]}-\epsilon^{\psi\gamma}_{\beta}\epsilon_{\chi\delta\alpha}\textsf{E}^{\beta\alpha},\\\nonumber
R^{\psi\gamma}_{(IV)\chi\delta}
&=&2(R_{m\chi}R^{m\psi}\delta^{\gamma}_{\delta}-R_{m\delta}R^{m\psi}\delta^{\gamma}_{\chi}+R_{m\chi}R^{m\gamma}
\delta^{\psi}_{\delta}+R_{m\delta}R^{m\gamma}\delta^{\psi}_{\chi})f_{G}\\\nonumber&+&2R^{ml}(R^{\psi}_{m\chi
l}\delta^{\gamma}_{\delta} -R^{\psi}_{m\delta
l}\delta^{\gamma}_{\chi}-R^{\gamma}_{m\chi
l}\delta^{\psi}_{\delta}+R^{\gamma}_{m\delta
l}\delta^{\psi}_{\chi})f_{G}-
R(R^{\psi}_{\chi}\delta^{\gamma}_{\delta}\\\nonumber&-&R^{\psi}_{\delta}\delta^{\gamma}_{\chi}-R^{\gamma}_{\chi}\delta^{\psi}_{\gamma}
+R^{\gamma}_{\delta}\delta^{\psi}_{\chi})f_{G}-2(R^{\psi}_{m\chi
l}\delta^{\gamma}_{\delta}-R^{\psi}_{m\delta
l}\delta^{\gamma}_{\chi}-R^{\gamma}_{m\chi
l}\delta^{\psi}_{\delta}\\\nonumber&+&R^{\gamma}_{m\delta
l}\delta^{\psi}_{\chi})\nabla^{m}\nabla^{l}f_{G}+2(R^{\psi}_{\chi}\delta^{\gamma}_{\delta}-R^{\psi}_{\delta}\delta^{\gamma}_{\chi}
-R^{\gamma}_{\chi}\delta^{\psi}_{\gamma}
+R^{\gamma}_{\delta}\delta^{\psi}_{\chi})\Box
f_{G}\\\nonumber&+&R(\delta^{\gamma}_{\delta}\nabla^{\psi}_{\chi}-\delta^{\gamma}_{\chi}\nabla^{\psi}_{\delta}
-\delta^{\psi}_{\delta}\nabla^{\gamma}_{\chi}+\delta^{\psi}_{\chi}\nabla^{\gamma}_{\delta})f_{G}-2(R^{m\psi}\delta^{\gamma}_{\delta}
\nabla_{\chi}\nabla_{m}\\\nonumber&-&R^{m\psi}\delta^{\gamma}_{\chi}\nabla_{\delta}\nabla_{m}-R^{m\gamma}\delta^{\psi}_{\delta}\nabla_{\chi}\nabla_{m}
+R^{m\gamma}\delta^{\psi}_{\chi}\nabla_{\delta}\nabla_{m})f_{G}-2(R^{m}_{\chi}\\\nonumber&\times&\delta^{\gamma}_{\delta}\nabla^{\psi}\nabla_{m}
-R^{m}_{\delta}\delta^{\gamma}_{\chi}\nabla^{\psi}\nabla_{m}
-R^{m}_{\chi}\delta^{\psi}_{\delta}\nabla^{\gamma}\nabla_{m}+R^{m}_{\delta}\delta^{\psi}_{\chi}\nabla^{\gamma}\nabla_{m})f_{G}
\\\nonumber&-&(R_{\chi lmn}R^{lmn\psi}\delta^{\gamma}_{\delta}-R_{\delta lmn}R^{lmn\psi}
\delta^{\gamma}_{\chi}-R_{\chi lmn}R^{lmn
\gamma}\delta^{\psi}_{\delta}+R_{\delta
lmn}\\\nonumber&\times&R^{lmn \gamma}\delta^{\psi}_{\chi})f_{G}
+\frac{1}{3}\left[(\varrho+p)f_{T}+2f+4R_{l\alpha}R^{m\alpha}f_{G}+4R^{lm}\right.\\\nonumber&\times&R^{\alpha}_{l\alpha
m}f_{G}
-2R^{\beta}_{lmn}R^{lmn}_{\alpha}f_{G}-4R^{m\beta}\nabla_{m}\nabla_{\beta}f_{G}-4R^{\alpha}_{l\alpha
m}\nabla^{m}\nabla^{l}f_{G}\\\nonumber&-&\left.2R\Box
f_{G}+16R^{lm}\nabla_{m}\nabla_{l}f_{G}
-4R^{m\alpha}\nabla_{\alpha}\nabla_{l}f_{G}
-2R^{2}f_{G}\right]\\\nonumber&-&\frac{\textsf{q}^2}{6r^4}h^{[\psi}_{[\chi}
\delta^{\gamma]}_{\delta]}+\frac{\textsf{q}^2}{2r^4}\mathrm{u}^{[\psi}\mathrm{u}_{[\chi}\delta^{\gamma]}_{\delta]}
-\frac{\textsf{q}^2}{r^4}\left(\phi^{[\psi}\phi_{[\chi}\delta^{\gamma]}_{\delta]}+\frac{1}{3}h^{[\psi}_{[\chi}\delta^{\gamma]}_{\delta]}\right)
\\\label{30a}&+&\frac{\textsf{q}^2}{12\pi
r^4}\mathrm{u}^{[\psi}\mathrm{u}_{[\chi}\delta^{\gamma]}_{\delta]}-\frac{\textsf{q}^2}{8\pi
r^4}\delta^{[\psi}_{[\chi}\delta^{\gamma]}_{\delta]},
\end{eqnarray}
we have employed the following formulas
\begin{equation}\nonumber
\epsilon_{\psi\gamma\chi}=\mathrm{u}^{\beta}\eta_{\beta\psi\gamma\chi},
\quad \epsilon_{\psi\gamma\chi}\mathrm{u}^{\chi}=0,
\end{equation}
\begin{equation}\nonumber
\epsilon^{\beta\gamma\alpha}\epsilon_{\alpha\psi\chi}=\delta^{\beta}_{\psi}h^{\gamma}_{\chi}-\delta^{\beta}_{\psi}h^{\gamma}_{\chi}
+\mathrm{u}_{\psi}(\mathrm{u}^{\beta}\delta^{\gamma}_{\chi}-\delta^{\beta}_{\chi}\mathrm{u}^{\gamma}).
\end{equation}
The following tensors in terms of fluid variables can be written as
\begin{eqnarray}\nonumber
\textsf{Y}_{\psi\chi} &=&
\frac{1}{6}(\varrho+p)h_{\psi\chi}f_{T}+\frac{\textsf{q}^2
h_{\psi\chi}}{3 r^4}+\frac{4\pi}{3}
(\varrho+3p)h_{\psi\chi}+\textsf{E}_{\psi\chi}\\\label{36a}&-&4\pi
\Pi_{\psi\chi}-\frac{\textsf{q}^2 }{
r^4}({\phi_{\psi}\phi_{\chi}-\frac{1}{3}h_{\psi\chi}})+\frac{\Pi_{\psi\chi}}{2}f_{T}
+\textsf{D}^{GT}_{\psi\chi},
\\\nonumber
\textsf{X}_{\psi\chi}&=& \frac{8\pi \varrho
h_{\psi\chi}}{3}+\frac{\textsf{q}^2 h_{\psi\chi}}{3
r^4}+\frac{\Pi_{\psi\chi}}{2}f_{T}-4\pi
\Pi_{\psi\chi}-\frac{\textsf{q}^2 }{
r^4}({\phi_{\psi}\phi_{\chi}-\frac{1}{3}h_{\psi\chi}})\\\label{31}&-&\textsf{E}_{\psi\chi}+\textsf{M}^{GT}_{\psi\chi},\\\label{32}
\textsf{Z}_{\psi\chi}&=&\textsf{N}^{GT}_{\psi\chi}.
\end{eqnarray}
The extra curvature terms $\textsf{D}^{GT}_{\psi\chi}$,
$\textsf{N}^{GT}_{\psi\chi}$ and $\textsf{M}^{GT}_{\psi\chi}$
present in the above equations are displayed in Appendix \textbf{A}.

The tensors $\textsf{X}_{\psi\chi}$ and $\textsf{Y}_{\psi\chi}$ are
subdivided into its trace and trace-free parts, which are denoted by
$\textsf{X}_{\textsf{T}},\textsf{X}_{\textsf{TF}},\textsf{Y}_{\textsf{T}},\textsf{Y}_{\textsf{TF}}$,
respectively. For the charged static cylindrical system, the scalar
corresponding to the tensor $\textsf{Z}_{\psi\chi}$ disappears. The
obtained scalars are represented as
\begin{eqnarray}\label{33}
\textsf{X}_{\textsf{T}}&=& 8\pi\big(\varrho+\frac{\textsf{q}^2}{8\pi
r^4}\big)+\textsf{O}^{GT},
\\\label{34} \textsf{X}_{\textsf{TF}} &=& -4\pi
\Pi-\frac{\textsf{q}^2}{
r^4}+\frac{\Pi}{2}f_{T}-\textsf{E},\\\label{35}
\textsf{Y}_{\textsf{T}} &=& \frac{\textsf{q}^2}{
r^4}+\frac{\left(\varrho+p\right)f_{T}}{2}+4\pi\left(\varrho+3p_{r}-2\Pi
\right)+\textsf{F}^{GT},
\\\label{36} \textsf{Y}_{\textsf{TF}} &=& \textsf{E}-\frac{\textsf{q}^2}{
r^4}+\frac{\Pi}{2}f_{T}-4\pi \Pi+\textsf{I}^{GT},
\end{eqnarray}
where
$\textsf{I}^{GT}=\frac{\textsf{Q}^{GT}_{\psi\chi}}{\phi_{\psi}\phi_{\chi}-\frac{1}{3}h_{\psi\chi}}$
and the terms $\textsf{O}^{GT}$, $\textsf{F}^{GT}$ and
$\textsf{Q}^{GT}$ are given in Appendix \textbf{A}. The scalar
$\textsf{X}_{\textsf{T}}$ is associated with the contribution of
energy density, whereas $\textsf{Y}_{\textsf{T}}$ controls the
anisotropic stresses in the presence of correction terms and total
charge of the system. The scalars $\textsf{X}_{\textsf{TF}}$ and
$\textsf{Y}_{\textsf{TF}}$ are rewritten by making use of
Eq.(\ref{22}) as
\begin{eqnarray}\label{37}
\textsf{X}_{\textsf{TF}} &=&
\frac{4\pi}{r^{3}}\int_{0}^{r}r^{3}(\varrho^{\textsf{eff}}-\frac{\textsf{q}^2}{8\pi
r^4})^{'}dr-\frac{3\textsf{q}^2}{2 r^4}+\frac{\Pi}{2}f_{T}-4\pi\Pi
^{GT}, \\\label{38} \textsf{Y}_{\textsf{TF}}&=&8\pi
\Pi+\frac{\Pi}{2}f_{T}-\frac{4\pi}{r^{3}}\int_{0}^{r}r^{3}\big(\varrho^{\textsf{eff}}-\frac{\textsf{q}^2}{8\pi
r^4}\big)^{'}dr+\textsf{I}^{GT}+4\pi\Pi^{GT}.
\end{eqnarray}
It should be noted that $\textsf{X}_{\textsf{TF}}$ gives the effects
of inhomogeneity caused by the energy density, charge and modified
terms. On the other hand, the scalar
$\textsf{Y}_{\textsf{\textsf{TF}}}$ includes the pressure
anisotropy, inhomogeneous energy density, charge and correction
terms. The complexity factor for a charged static cylindrical
configuration is based on the assumption that the factor must
incorporate all basic physical variables, electromagnetic effects
and extra curvature terms. As a result,
$\textsf{Y}_{\textsf{\textsf{TF}}}$ is chosen as the complexity
factor. Using Eq.(\ref{38}) in (\ref{27}) to discuss physical
meaning of $\textsf{Y}_{\textsf{\textsf{TF}}}$
\begin{equation}\label{39}
m_{T}=(m_{T})_{\Sigma}\left(\frac{r}{r_{\Sigma}}\right)^{3}+r^{3}\int^{r_{\Sigma}}_{r}\frac
{KL}{r} [ \textsf{Y}_{\textsf{TF}}-\frac{3\textsf{q}^2}{2
r^4}-\frac{\Pi}{2}f_{T}-\textsf{I}^{GT}+4\pi \Pi^{GT}]dr.
\end{equation}
When we compare Eqs.(\ref{27}) and (\ref{39}), it can be observed
that Tolman mass is affected by anisotropic pressure, charge terms
and inhomogeneous energy density due to
$\textsf{Y}_{\textsf{\textsf{TF}}}$.

\section{Zero Complexity Factor Condition}

The behavior of celestial objects is governed by a vast array of
interconnected physical factors. The least complicated structures
are those that possess isotropic properties, like a cylinder
occupying only dust. However, the factors responsible for generating
complexity in the current setup are pressure anisotropy, charge
terms, inhomogeneity of energy density and the presence of modified
terms. The structure scalar $\textsf{Y}_{\textsf{TF}}$ is taken as
the complexity factor on the basis of the aforementioned parameters.
It also describes how these factors affect the Tolman mass.
Substituting $\textsf{Y}_{\textsf{TF}}=0$, we set the zero
complexity factor constraint as
\begin{equation}\label{40}
\Pi=\frac{1}{4\pi+\frac{f_{T}}{2}}\left[\frac{4\pi}{r^{3}}\int^{r}_{0}r^{3}\big(\varrho^{\textsf{eff}}-\frac{\textsf{q}^2}{8\pi
r^4}\big)^{'}dr+\frac{\textsf{q}^2}{ 2r^4}-I^{GT}-4\pi
\Pi^{GT}\right].
\end{equation}
In order to construct solution to the modified field equations, this
condition will be utilized as an extra limitation.

\subsection{Gokhroo-Mehra Model}

Gokhroo and Mehra \cite{25a} addressed a particular form of the
energy density to evaluate the solution of the modified field
equations as well as the behavior of anisotropic stellar system.
They designed a system that describes both the dynamics of neutron
stars as well as larger redshifts of numerous quasi-stellar
configurations. For the considered setup, the variable energy
density is
\begin{equation}\label{41}
\varrho^{\textsf{eff}}=\varrho_{o}\left(1-\frac{\textsf{k}r^{2}}{r^{2}_{\Sigma}}\right).
\end{equation}
After making use of this expression in Eq.(\ref{15}), the mass
function will become
\begin{equation}\label{42}
m=\frac{\textsf{q}^2}{2r}+\frac{r}{8}-\zeta
r^3\left(\frac{\textsf{k}r^{2}}{5r^{2}_{\Sigma}}-\frac{1}{3}\right)+4\pi
\int\frac{\textsf{q}^2}{8\pi r^2},
\end{equation}
where $\textsf{k}$ is a constant in the range $(0,1)$ and
$\zeta=4\pi \varrho_{o}$. When we compare Eqs.(\ref{15}) and
(\ref{42}), the expression for the metric function becomes
\begin{equation}\label{43}
L=\frac{1}{\sqrt {2\zeta
r^2\left(-\frac{1}{3}+\frac{\textsf{k}r^{2}}{5r^{2}_{\Sigma}}\right)+\frac{1}{r}
\int\frac{\textsf{q}^2}{r^2}}}.
\end{equation}
The field equations can also be composed in the following manner
\begin{equation}\label{44}
8\pi(p^{\textsf{eff}}_{r}-\frac{\textsf{q}^2}{8\pi
r^4}-p^{\textsf{eff}}_{\bot}-\frac{\textsf{q}^2}{8\pi
r^4})=\frac{1}{L^2}\left[\frac{K'}{rK}+\frac{1}{r^{2}}-\frac{K''}{K}+\frac{K'L'}{KL}
+\frac{L'}{rL}\right].
\end{equation}
Here, we introduce two new variables ($\textsf{y}$ and $\textsf{z}$)
\cite{12} as
\begin{equation}\label{45}
L^{-2}=\textsf{y}(r),\quad K^2=e^{\int(2\textsf{z}(r)-2/r)dr},
\end{equation}
and Eq.\eqref{44} in terms of $\textsf{y}$ and $\textsf{z}$ is given
as
\begin{equation}\label{46}
\textsf{y}'+\textsf{y}\left[-\frac{6}{r}+\frac{4}{\textsf{z}r^{2}}+2\textsf{z}+\frac{2\textsf{z}'}{\textsf{z}}\right]
=\frac{16\pi(\Pi^{\textsf{eff}}-\frac{\textsf{q}^2}{4\pi
r^4})}{\textsf{z}}.
\end{equation}
Its solution provides the metric functions ($K^2$ and $L^2$), thus,
the respective line element turns out to be
\begin{eqnarray}\nonumber
ds^{2}&=&-e^{\int(2\textsf{z}-2/r)dr}
dt^{2}+\frac{\textsf{z}^{2}e^{\int(2\textsf{z}(r)-\frac{4}{r^{2}\textsf{z}(r)})dr}}{r^{6}\left(16\pi\int\frac{\textsf{z}
(\Pi^{\textsf{eff}}-\frac{\textsf{q}^2}{4\pi
r^4})e^{\int(2\textsf{z}(r)-\frac{4}
{r^{2}\textsf{z}(r)})dr}}{r^6}dr+\textsf{H}\right)}dr^{2}\\\label{47}&+&r^{2}(d\theta^{2}+\alpha^2dz^2),
\end{eqnarray}
where the integration constant is denoted by $\textsf{H}$. The field
equations are rewritten in the light of new variables as
\begin{align}\label{48}
4\pi (p^{\textsf{eff}}_{r}-\frac{\textsf{q}^2}{8\pi
r^4})&=\left(\frac{1}{4}+\frac{\textsf{q}^2}{r^2}-\frac{2m}{r}\right)\left(\frac{\textsf{z}}{r}-\frac{1}{2r^2}\right),\\\label{49}
4\pi (\varrho^{\textsf{eff}}+\frac{\textsf{q}^2}{8\pi
r^4})&=\frac{m'}{r^{2}}-\frac{\textsf{q}\textsf{q}^{'}}{
r^3}+\frac{\textsf{q}^2}{2r^4}-\frac{1}{8r^2},\\\nonumber 8\pi
(p^{\textsf{eff}}_{\bot}+\frac{\textsf{q}^2}{8\pi
r^4})&=\bigg(\frac{1}{4}+\frac{\textsf{q}^2}{
r^2}-\frac{2m}{r}\bigg)\bigg(\textsf{z}'+\textsf{z}^{2}-\textsf{z}\left(\frac{4rm'+\frac{\textsf{4q}^2}{
r}-{4\textsf{q}\textsf{q}^{'}}-4m}{r^2-8mr+4\textsf{q}^2}\right)\\\label{50}&-\frac{\textsf{z}}{r}+\frac{1}{r^{2}}\bigg).
\end{align}

\subsection{Polytrope with Zero Complexity Constraint}

The interior regime of self-gravitating systems is determined by a
variety of physical factors (density, temperature, pressure etc.),
each of which plays a specific function.  However, the contribution
of some variables over others is more important in analyzing the
structure. In this regard, it is beneficial to possess an equation
of state that efficiently investigates the behavior of anisotropic
self-gravitating objects. Arias et al. \cite{27} discussed three
anisotropic models using zero complexity condition (as a constraint)
to close the system. They also illustrated the graphical analysis of
polytropic system. Khan et al. \cite{27a} studied the generalized
polytropes for charged spherical and cylindrical geometries and
solved them through the complexity factor condition. The anisotropic
polytropic equation of state (relation between energy density and
pressure) is given as \cite{26a}
\begin{equation}\label{51}
p^{\textsf{eff}}_{r}=\varrho^{\textsf{eff}(\omega)}=B\varrho^{\textsf{eff}(1+1/n)},
\end{equation}
where $B$ stands for the polytropic constant, $\omega$ denotes the
polytropic exponent and $n$ represents the polytropic index. Here,
we utilize the linear $f(G,T)$ gravity model \cite{28} and known
expression for charge \cite{28a} to describe the interior
configuration, respectively, as
\begin{align}\label{56}
f(G,T)= G+\varepsilon T, \quad \textsf{q}=\zeta r^3.
\end{align}
Introducing the dimensionless quantities, we have
\begin{eqnarray*}\label{52}
\sigma&=&\frac{p^{\textsf{eff}}_{rc}}{\varrho^{\textsf{eff}}_{c}},
\quad r=\frac{\xi}{J},\quad
J^{2}=\frac{4\pi\varrho^{\textsf{eff}}_{c}}{\sigma(1+n)},\\
\Psi^{n}&=&\frac{\varrho^{\textsf{eff}}}{\varrho^{\textsf{eff}}_{c}},
\quad \vartheta(\xi)=m(r)J^{3}/4\pi\varrho^{\textsf{eff}}_{c}.
\end{eqnarray*}
Substituting these variables in Eqs.(\ref{11}) and (\ref{15}), we
obtain the Tolman-Oppenheimer-Volkoff equation and mass function as
\begin{eqnarray}\nonumber
&&\xi^{2}\frac{d\Psi}{d\xi}\left[\frac{1+\frac{4\textsf{q}^2J^{2}}{\xi^{2}}-8\vartheta
\sigma
(n+1)/\xi}{1+\sigma\Psi}\right]-4\left\{\xi^{3}(-\sigma\Psi^{n+1}-\frac{\textsf{q}^2J^4}{\varrho^{\textsf{eff}}8\pi
\xi^4})-\vartheta\right\}\\\nonumber&-&\frac{\xi}{2(n+1)\sigma}-\frac{4\textsf{q}^2J^{2}}{2\sigma
(n+1)\xi}+\frac{2(\Pi-\frac{\textsf{q}^2 J^4}{4\pi
\xi^4})\xi\Psi^{-n}}
{p^{\textsf{eff}}_{rc}(n+1)}\left[\frac{1+\frac{4\textsf{q}^2J^{2}}{\xi^{2}}-8\vartheta
\sigma(n+1)/\xi}{1+\sigma\Psi}\right]\\\label{53}
&=&\xi^{2}\left[\frac{1+\frac{4\textsf{q}^2J^{2}}{\xi^{2}}-8\vartheta
\sigma
(n+1)/\xi}{\Psi^{n}p^{\textsf{eff}}_{rc}(1+\sigma\Psi)(n+1)J}\right]\bigg[\textsf{Z}L^2+\bigg(\frac{\textsf{q}^2J^4}{8\pi
\xi^4}\bigg)^{'}\bigg],
\end{eqnarray}
\begin{equation}\label{54}
\frac{d\vartheta}{d\xi}=\frac{1}{2\sigma
(n+1)}\bigg[\frac{\textsf{q}^2J^{2}}{\xi^{2}}+\frac{J^{2}}{4\pi}+\bigg(\frac{\textsf{q}^2J}{\xi}\bigg)^{'}\bigg]+\Psi^{n}\xi^{2}.
\end{equation}
where the evaluation of these quantities at the center is
represented by subscript $c$. These ordinary differential equations
are composed of three unknowns, i.e., $\Psi, \vartheta$ and $\Pi$
that further depend on $\sigma$ and $n$.  It is noticed that the
number of unknowns are greater than the number of equations,
therefore, we require one more equation to obtain a unique solution.
For this reason, the complexity free condition can be used as a
constraint and takes the form
 \begin{eqnarray}\nonumber
\frac{3\Pi}{\xi}+\frac{d\Pi}{d\xi}&=&
\frac{1}{(4\pi+\frac{\varepsilon}{2})}\bigg[4\pi\varrho^{\textsf{eff}}_{c}n\Psi^{n-1}\frac{d\Psi}{d\xi}
+\frac{\textsf{q}\textsf{q}^{'}J^3}{4\pi\xi^4}-\frac{3}{\xi}\bigg(4\pi\Pi^{GT}+I^{GT}\bigg)\\\label{55}
&-&\frac{\textsf{q}^2J^4}{\xi^5} -4\pi
\frac{d\Pi^{\textsf{eff}}}{d\xi}+\frac{\textsf{q}\textsf{q}^{'}J^3}{\xi^4}
-\frac{\textsf{q}^2J^4}{2\pi\xi^5}-I_{, \xi}^{GT}\bigg].
\end{eqnarray}
Thus a system of three equations (\ref{53})-(\ref{55}) is obtained
that will represent a unique solution. Using this solution for
certain values of $\sigma$ and $n$, it is possible to study physical
characteristics of quantities, such as density and mass. Thus, to
illustrate the graphical analysis of dimensionless quantities (mass,
density and anisotropy), we select the values of unknowns as
$\sigma=0.1$, $\varepsilon=0.5$, $\zeta=0.001$, $n=0.2$ and
$\varrho^{\textsf{eff}}_{c}=2$. Figure \textbf{1} depicts the
behavior of density, mass function and anisotropy, respectively. It
can be observed that density is maximum at the center while goes on
decreasing with increasing $\xi$. In the 2nd plot, the mass function
exhibits the increasing trend for higher values of $\xi$. On the
other hand, the last plot indicates that the anisotropy of the
system is zero closer to the center and follows a decreasing
behavior as $\xi$ increases.
\begin{figure}\center
\epsfig{file=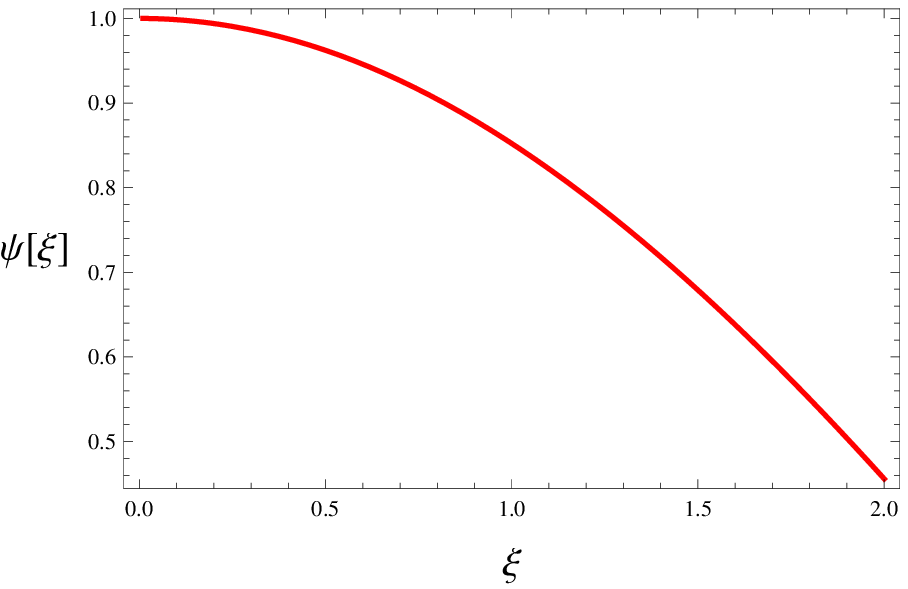,width=0.4\linewidth}\epsfig{file=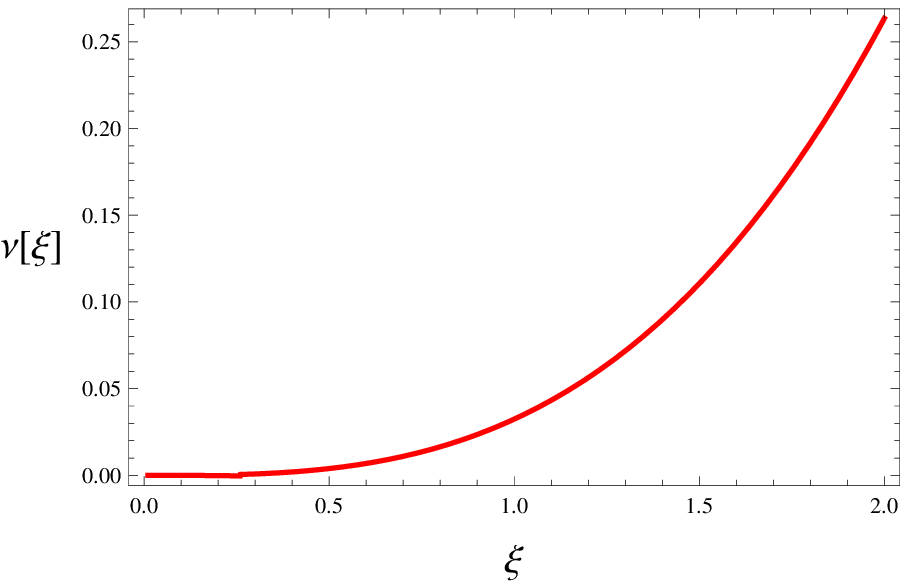,width=0.4\linewidth}
\epsfig{file=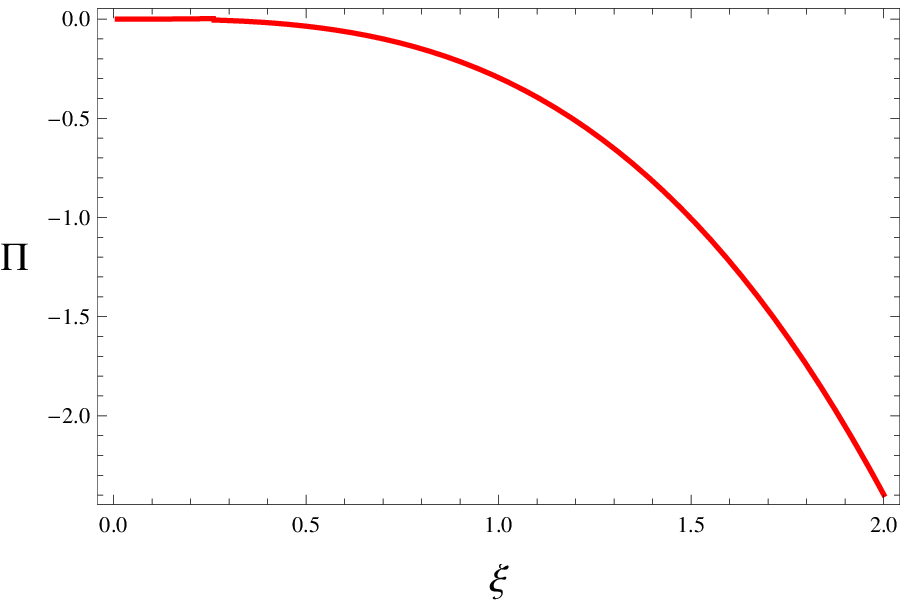,width=0.4\linewidth}\caption{Behavior of
density, mass function and anisotropy versus $\xi$}
\end{figure}

\section{Concluding Remarks}

In this paper, we analyze the influence of electromagnetic field to
gauge the complexity factor of static cylindrical symmetric
spacetime within the formalism of $f(G,T)$ theory. We have
established the charged modified field equations together with
hydrostatic equilibrium condition corresponding to anisotropic
matter source. We have calculated mass function of the system using
C-energy formalism and Tolman mass formula. We have also discussed
relationship between the mass functions with physical variables as
well as the Weyl tensor. A system that has isotropic pressure and
homogenous energy density is thought to be less complicated. In
other words, when both terms cancel the influence of each other, the
system is said to be complexity free. The splitting of the Riemann
tensor yields four structure scalars
$\textsf{X}_{\textsf{\textsf{T}}}$,$\textsf{X}_{\textsf{\textsf{TF}}}$,
$\textsf{Y}_{\textsf{\textsf{T}}}$,$\textsf{Y}_{\textsf{\textsf{TF}}}$)
and we have followed Herrera technique to determine the complexity
factor. The reasons to recognize the scalar function
$\textsf{Y}_{\textsf{\textsf{TF}}}$ as complexity factor are given
below.
\begin{enumerate}
\item All physical parameters such as energy density inhomogeneity
and pressure anisotropy along with extra curvature terms are
incorporated in this factor.
\item This scalar depicts the influence of pressure anisotropy and inhomogeneous
source on the Tolman mass.
\item It also encompasses the electromagnetic effects for
the charged cylindrical distribution.
\end{enumerate}

In the current setup, the existence of homogeneous energy density
and isotropic pressure does not correspond to a complex free
structure as a result of correction terms (opposite to GR).
Consequently, the additional curvature terms cause to increase the
complexity of the cylindrical configuration. It is worthy to mention
here that the absence of correction terms and charges along with
homogeneous energy density and isotropic pressure leads to
complexity free system. To achieve a complex free system, we have
substituted $\textsf{Y}_{\textsf{\textsf{TF}}}=0$ which gives rise
to vanishing complexity condition that contains additional terms. We
have used this constraint to find solution for the two cases.

In the first case, Gokhroo-Mehra model is used to examine physical
aspects of compact entities. It is found that if specific values are
assigned to $m,z$ this model will correspond to distinct solution.
For the second case, we have considered the polytropic equation of
state and have introduced some dimensionless quantities. This leads
to form a system of three dimensionless equations (mass,
Tolman-Oppenheimer-Volkoff equation and zero complexity condition).
For this model, we have investigated the graphical behavior of the
system through numerical approach by assigning some particular
values to the involved unknowns. It is seen that the system
represents a decline in energy density. whereas mass enlarges for
larger $\xi$. However, the system has zero anisotropy at the center
which becomes negative as $\xi$ increases.

\section*{Appendix A}
\renewcommand{\theequation}{A\arabic{equation}}
\setcounter{equation}{0} The modified term \textsf{Z} is expressed
as
\begin{eqnarray}\nonumber
\textsf{Z}&=&\frac{f_{T}}{k^{2}-f_{T}}\left[\left(-p_{r}-p\right)\frac{f'_{T}}{f_{T}}-\frac{\textsf{q}\textsf{q}^{'}}{4\pi
r^4}+\left(-2p_{r}L^2-pL^2\right)'-\frac{L^2}{2}(-\varrho+3p)^{'}
\right].
\end{eqnarray}
The following describes the impact of additional terms in the scalar
structure
\begin{eqnarray}\nonumber
\textsf{N}^{GT}_{\psi\chi}&=&\left[2R_{m\delta}R^{m\gamma}f_{G}-RR^{\gamma}_{\delta}f_{G}
+2R^{lm}R^{\gamma}_{l\delta m}f_{G}-2R_{\delta
lmn}R^{lmn\gamma}f_{G}\right.\\\nonumber
&+&\left.2R^{\gamma}_{\delta}\Box
f_{G}-2R^{m\gamma}\nabla_{\delta}\nabla_{m}f_{G}+R\nabla^{\gamma}\nabla_{\delta}f_{G}
-2R^{m}_{\delta}\nabla^{\gamma}\nabla_{m}f_{G}\right.\\\nonumber
&-&\left.2R^{\gamma}_{l\delta
m}\nabla^{m}\nabla^{l}f_{G}\right]\epsilon_{\psi\chi\gamma}\mathrm{u}^{\delta},
\\\nonumber
\textsf{D}^{GT}_{\psi\chi} &=&
2\left[R_{m\chi}R^{m}_{\psi}f_{G}-\frac{1}{2}RR_{\psi\chi}f_{G}+
R^{lm}R_{l\chi m\psi}f_{G}-\frac{1}{2}R_{\chi
lmn}R^{lmn}_{\psi}f_{G}+R_{\psi\chi}\Box
f_{G}\right.\\\nonumber&+&\left.\frac{1}{2}R
\nabla_{\psi}\nabla_{\chi}f_{G}-R^{m}_{\psi}\nabla_{\chi}\nabla_{m}f_{G}-R^{m}_{\chi}\nabla_{\psi}\nabla_{m}f_{G}-R_{l\chi
m\psi}\nabla^{m}\nabla^{l}f_{G}\right]\\\nonumber
&+&4R^{lm}h_{\psi\chi}\nabla_{m}\nabla_{l}f_{G}-2Rh_{\psi\chi}\Box
f_{G}+2\left[-R_{m\delta}R^{m}_{\psi}f_{G}- R^{lm}R_{l\delta
m\psi}f_{G}\right.\\\nonumber&+&\left.\frac{1}{2}R_{\delta
lmn}R^{lmn}_{\psi}f_{G}+\frac{1}{2}RR_{\psi\chi}f_{G}-R_{\delta\psi}\Box
f_{G}+R_{l\delta
m\psi}\nabla^{m}\nabla^{m}f_{G}\right.\\\nonumber&-&\left.\frac{1}{2}R
\nabla_{\psi}\nabla_{\delta}f_{G}+R^{m}_{\psi}\nabla_{\delta}\nabla_{m}f_{G}
+R^{m}_{\delta}\nabla_{\psi}\nabla_{m}f_{G}\right]\mathrm{u}_{\chi}\mathrm{u}^{\delta}+2\left[-R_{m\chi}R^{m\gamma}f_{G}\right.\\\nonumber&+&\left.\frac{1}{2}R_{\chi
lmn}R^{lmn\gamma}f_{G}- R^{lm}R^{\gamma}_{l\chi
m}f_{G}+\frac{1}{2}RR^{\gamma}_{\chi}f_{G}-R^{\gamma}_{\chi}\Box
f_{G}+R^{m\gamma}\nabla_{\chi}\nabla_{m}f_{G}\right.\\\nonumber&-&\left.\frac{1}{2}R
\nabla^{\gamma}\nabla_{\chi}f_{G}+R^{m}_{\chi}\nabla^{\gamma}\nabla_{m}f_{G}+R^{\gamma}_{m\chi
l}\nabla^{m}\nabla^{l}\right]\mathrm{u}_{\psi}\mathrm{u}_{\gamma}+
2\left[R_{m\delta}R^{m\gamma}f_{G}\right.\\\nonumber
&-&\left.\frac{1}{2}R_{\delta lmn}R^{lmn
\gamma}f_{G}+R^{lm}R^{\gamma}_{l\delta
m}f_{G}-\frac{1}{2}RR^{\gamma}_{\delta}f_{G}+R^{\gamma}_{\delta}\Box
f_{G}+\frac{1}{2}R
\nabla^{\gamma}\nabla_{\delta}f_{G}\right.\\\nonumber&-&\left.R^{\gamma}_{l\delta
m}\nabla^{m}\nabla^{l}f_{G}-R^{m\gamma}\nabla_{\delta}\nabla_{m}f_{G}-R^{m}_{\delta}\nabla^{\gamma}\nabla_{m}
f_{G}\right] g_{\psi\chi}\mathrm{u}_{\gamma}\mathrm{u}^{\delta}
-\frac{1}{3}\left[-2R^{2}f_{G}\right.\\\nonumber&-&\left.2R^{\beta}_{lmn}R^{lmn}_{\beta}f_{G}
-2R\Box
f_{G}+4R^{m\alpha}R_{m\alpha}f_{G}+4R^{lm}R^{\alpha}_{m\alpha
l}f_{G}+16R^{lm}\nabla_{m}\nabla_{l}f_{G}\right.\\\nonumber&-&\left.4R^{m\beta}\nabla_{\beta}\nabla_{m}f_{G}
-4R^{m\alpha}\nabla_{\alpha}\nabla_{m}f_{G}-4R^{\alpha}_{l\alpha
m}\nabla^{m}\nabla^{l}f_{G}\right]h_{\psi\chi}
-\frac{1}{6}fh_{\psi\chi},\\\nonumber \textsf{F}^{GT}&=&
2\left[R_{m\chi}R^{m}_{\psi}f_{G}+ R^{lm}R_{m\chi
l\psi}f_{G}-\frac{1}{2}RR_{\psi\chi}f_{G}-\frac{1}{2}R_{\chi\chi
lmn}R^{lmn}_{\psi}f_{G}+R_{\psi\chi}\Box
f_{G}\right.\\\nonumber&+&\left.\frac{1}{2}R
\nabla_{\psi}\nabla_{\chi}f_{G}-R^{m}_{\psi}\nabla_{\chi}\nabla_{m}f_{G}-R^{m}_{\chi}\nabla_{\psi}\nabla_{m}f_{G}-R_{m\chi
l\psi}\nabla^{m}\nabla^{l}f_{G}\right]g^{\psi\chi}\\\nonumber
&+&12R^{lm}\nabla_{m}\nabla_{l}f_{G}-6R\Box
f_{G}+2\left[-R_{m\delta}R^{m}_{\psi}f_{G}- R^{lm}R_{m\delta
l\psi}f_{G}+\frac{1}{2}RR_{\psi\delta}f_{G}\right.\\\nonumber&-&\left.\frac{1}{2}R
\nabla_{\psi}\nabla_{\delta}f_{G}-R_{\delta\psi}\Box
f_{G}+R_{m\delta
l\psi}\nabla^{m}\nabla^{l}f_{G}+R^{m}_{\psi}\nabla_{\delta}\nabla_{m}f_{G}
+R^{m}_{\delta}\nabla_{\psi}\nabla_{m}f_{G}\right.\\\nonumber&+&\left.\frac{1}{2}R_{\delta
lmn}R^{lmn}_{\psi}f_{G}\right]\mathrm{u}_{\chi}\mathrm{u}^{\delta}g^{\psi\chi}+2\left[-R_{m\chi}R^{m\gamma}f_{G}-
R^{lm}R^{\gamma}_{m\chi
l}f_{G}+\frac{1}{2}RR^{\gamma}_{\chi}f_{G}\right.\\\nonumber&+&\left.\frac{1}{2}R_{\chi
lmn}R^{lmn\gamma}f_{G}-R^{\gamma}_{\chi}\Box
f_{G}+R^{m\gamma}\nabla_{\chi}\nabla_{m}f_{G}+R^{m}_{\chi}\nabla^{\gamma}\nabla_{m}f_{G}+R^{\gamma}_{m\chi
l}\nabla^{m}\nabla^{l}f_{G}\right.\\\nonumber&-&\left.\frac{1}{2}R
\nabla^{\gamma}\nabla_{\chi}f_{G}\right]\mathrm{u}_{\psi}\mathrm{u}_{\gamma}g^{\psi\chi}+
2\left[R_{m\delta}R^{m\gamma}f_{G}+ R^{lm}R^{\gamma}_{m\delta
l}f_{G}-R^{m\gamma}\nabla_{\delta}\nabla_{m}f_{G}\right.\\\nonumber
&-&\left.\frac{1}{2}R_{\delta lmn}R^{lmn
\gamma}f_{G}+R^{\gamma}_{\delta}\Box f_{G}+\frac{1}{2}R
\nabla^{\gamma}\nabla_{\delta}f_{G}-\frac{1}{2}RR^{\gamma}_{\delta}f_{G}-R^{m}_{\delta}\nabla^{\gamma}\nabla_{m}
f_{G}\right.\\\nonumber&-&\left.R^{\gamma}_{m\delta
l}\nabla^{m}\nabla^{l}f_{G}\right]
g_{\psi\chi}\mathrm{u}_{\gamma}\mathrm{u}^{\delta}g^{\psi\chi}
-\left[4R^{l\alpha}R_{l\alpha}f_{G}+4R^{lm}R^{\alpha}_{l\alpha
m}f_{G}-2R^{2}f_{G}\right.\\\nonumber&-&\left.2R^{\beta}_{lmn}R^{lmn}_{\beta}f_{G}-2R\Box
f_{G}+16R^{lm}\nabla_{m}\nabla_{l}f_{G}
-4R^{m\alpha}\nabla_{\alpha}\nabla_{m}f_{G}\right.\\\nonumber&-&\left.4R^{m\beta}\nabla_{\beta}\nabla_{m}f_{G}
-4R^{\alpha}_{l\alpha m}\nabla^{m}\nabla^{l}f_{G}\right]-
\frac{1}{2}f,
\end{eqnarray}
\begin{eqnarray}\nonumber
\textsf{Q}^{GT}_{(\psi\chi)} &=&
\left[2R_{md}R^{m}_{c}f_{G}+2R^{lm}R_{ldmc}f_{G}-RR_{cd}f_{G}-R_{dlmn}R^{lmn}_{c}f_{G}+2R_{cd}\Box
f_{G}\right.\\\nonumber&+&\left.R\nabla_{c}\nabla_{d}f_{G}-2R^{m}_{c}\nabla_{d}\nabla_{m}f_{G}-2R^{m}_{d}\nabla_{c}\nabla_{m}f_{G}
-2R_{ldmc}\nabla^{m}\nabla^{l}f_{G}\right]
h^{c}_{\psi}h^{d}_{\chi}\\\nonumber&+&2\left[R_{m\delta}R^{m\gamma}f_{G}+
R^{lm}R^{\gamma}_{m\delta
l}f_{G}-\frac{1}{2}RR^{\gamma}_{\delta}f_{G}-\frac{1}{2}R_{\delta
lmn}R^{lmn
\gamma}f_{G}\right.\\\nonumber&+&\left.R^{\gamma}_{\delta}\Box
f_{G}+\frac{1}{2}R
\nabla^{\gamma}\nabla_{\delta}f_{G}-R^{m\gamma}\nabla_{\delta}\nabla_{m}f_{G}
-R^{m}_{\delta}\nabla^{\gamma}\nabla_{m}f_{G}\right.\\\nonumber&-&\left.R^{\gamma}_{m\delta
l}\nabla^{m}\nabla^{l}f_{G}\right]h_{\psi\chi}\mathrm{u}_{\gamma}\mathrm{u}^{\delta}-2\left[R_{m\chi}R^{m}_{\psi}f_{G}+
R^{lm}R_{m\chi
l\psi}f_{G}-\frac{1}{2}RR_{\psi\chi}f_{G}\right.\\\nonumber&-&\left.\frac{1}{2}R_{\chi
lmn}R^{lmn}_{\psi}f_{G}+R_{\psi\chi}\Box f_{G}+\frac{1}{2}R
\nabla_{\psi}\nabla_{\chi}f_{G}-R^{m}_{\psi}\nabla_{\chi}\nabla_{m}f_{G}\right.\\\nonumber&-&\left.R^{m}_{\chi}\nabla_{\psi}\nabla_{m}f_{G}-R_{m\chi
l\psi}\nabla^{m}\nabla^{l}f_{G}\right]-2\left[-R_{m\delta}R^{m}_{\psi}f_{G}-
R^{lm}R_{m\delta
l\psi}f_{G}\right.\\\nonumber&+&\left.\frac{1}{2}RR_{\psi\delta}f_{G}+\frac{1}{2}R_{\delta
lmn}R^{lmn}_{\psi}f_{G}-R_{\delta\psi}\Box f_{G}+R_{m\delta
l\psi}\nabla^{m}\nabla^{l}f_{G}\right.\\\nonumber&+&\left.R^{m}_{\psi}\nabla_{\delta}\nabla_{m}f_{G}
+R^{m}_{\delta}\nabla_{\psi}\nabla_{m}f_{G}-\frac{1}{2}R
\nabla_{\psi}\nabla_{\delta}f_{G}\right]\mathrm{u}_{\chi}\mathrm{u}^{\delta}-2\left[-R_{m\chi}R^{m\gamma}f_{G}\right.\\\nonumber&-&\left.R^{lm}R^{\gamma}_{m\chi
l}f_{G}+\frac{1}{2}RR^{\gamma}_{\chi}f_{G}+\frac{1}{2}R_{\chi
lmn}R^{lmn\gamma}f_{G}-R^{\gamma}_{\chi}\Box
f_{G}\right.\\\nonumber&+&\left.R^{m\gamma}\nabla_{\chi}\nabla_{m}f_{G}+R^{m}_{\chi}\nabla^{\gamma}\nabla_{m}f_{G}+R^{\gamma}_{m\chi
l}\nabla^{m}\nabla^{l}f_{G}-\frac{1}{2}R
\nabla^{\gamma}\nabla_{\chi}f_{G}\right]\mathrm{u}_{\psi}\mathrm{u}_{\gamma}\\\nonumber&-&2\left[R_{m\delta}R^{m\gamma}f_{G}+
R^{lm}R^{\gamma}_{m\delta
l}f_{G}-\frac{1}{2}RR^{\gamma}_{\delta}f_{G}-\frac{1}{2}R_{\delta
lmn}R^{lmn
\gamma}f_{G}\right.\\\nonumber&+&\left.R^{\gamma}_{\delta}\Box
f_{G}+\frac{1}{2}R
\nabla^{\gamma}\nabla_{\delta}f_{G}-R^{m\gamma}\nabla_{\delta}\nabla_{m}f_{G}
-R^{m}_{\delta}\nabla^{\gamma}\nabla_{m}f_{G}\right.\\\nonumber&-&\left.R^{\gamma}_{m\delta
l}\nabla^{m}\nabla^{l}f_{G}\right]\mathrm{u}_{\gamma}\mathrm{u}^{\delta}g_{\psi\chi},
\end{eqnarray}
\begin{eqnarray}\nonumber
\textsf{M}^{GT}_{\psi\chi}  &=&\left[\frac{1}{2}R_{m\epsilon}R^{m
p}f_{G}+\frac{1}{2}R^{lm}R^{p}_{m\epsilon
l}f_{G}-\frac{1}{4}RR^{p}_{\epsilon}f_{G}-\frac{1}{4}R_{\epsilon
lmn}R^{lmn p}f_{G}\right.\\\nonumber&+&\left.\frac{1}{2}
R^{p}_{\epsilon}\Box
f_{G}+\frac{1}{4}R\nabla^{p}\nabla_{\epsilon}f_{G}-\frac{1}{4}R^{lp}\nabla_{\epsilon}\nabla_{m}f_{G}-\frac{1}{2}R^{m}_{\epsilon}\nabla^{p}\nabla_{m}
f_{G}\right.\\\nonumber&-&\left.\frac{1}{2}R^{p}_{m\epsilon
l}\nabla^{m}\nabla^{l}f_{G}\right]\epsilon_{p
\delta\chi}\epsilon^{\epsilon\delta}_{\psi}+\left[-\frac{1}{2}R_{m\delta}R^{lp}f_{G}-\frac{1}{2}R^{lm}R^{p}_{m\delta
l}f_{G}\right.\\\nonumber&+&\left.\frac{1}{4}RR^{p}_{\delta}f_{G}+\frac{1}{4}R_{\delta
lmn}R^{lmnp}f_{G}-\frac{1}{2}R^{p}_{\delta}\Box
f_{G}-\frac{1}{4}R\nabla^{p}\nabla_{\delta}f_{G}\right.\\\nonumber&+&\left.\frac{1}{4}R^{lp}\nabla_{\delta}\nabla_{m}f_{G}
+\frac{1}{2}R^{m}_{\delta}\nabla^{p}\nabla_{m}f_{G}
+\frac{1}{2}R^{p}_{m\delta
l}\nabla^{m}\nabla^{l}f_{G}\right]\epsilon_{p\epsilon
\chi}\epsilon^{\epsilon\delta}_{\psi}\\\nonumber&+&\left[-\frac{1}{2}R_{m\epsilon}R^{m\gamma}f_{G}-\frac{1}{2}R^{lm}R^{\gamma}_{m\epsilon
l}f_{G}
+\frac{1}{4}RR^{\gamma}_{\epsilon}f_{G}+\frac{1}{4}R_{\epsilon
lmn}R^{lmn\gamma}f_{G}\right.\\\nonumber&-&\left.\frac{1}{2}R^{\gamma}_{\epsilon}\Box
f_{G}-\frac{1}{4}R\nabla^{\gamma}\nabla_{\epsilon}f_{G}+\frac{1}{4}R^{m\gamma}\nabla_{\epsilon}\nabla_{m}f_{G}
+\frac{1}{2}R^{m}_{\epsilon}\nabla^{\gamma}\nabla_{m}
f_{G}\right.\\\nonumber&+&\left.\frac{1}{2}R^{\gamma}_{m\epsilon
l}\nabla^{m}
\nabla^{l}f_{G}\right]\epsilon_{\delta\gamma\chi}\epsilon^{\epsilon\delta}_{\psi}+\left[\frac{1}{2}R_{m\delta}R^{m\gamma}f_{G}
+\frac{1}{2}R^{lm}R^{\gamma}_{m\delta
l}f_{G}\right.\\\nonumber&-&\left.\frac{1}{4}RR^{\gamma}_{\delta}f_{G}-\frac{1}{4}R_{\delta
lmn}R^{lmn\gamma}f_{G}+\frac{1}{2}R^{\gamma}_{\delta}\Box
f_{G}+\frac{1}{4}R\nabla^{\gamma}\nabla_{\delta}f_{G}\right.\\\nonumber&-&\left.\frac{1}{4}R^{m\gamma}\nabla_{\delta}\nabla_{m}f_{G}
-\frac{1}{2}R^{m}_{\delta}\nabla^{\gamma}\nabla_{m}f_{G}-\frac{1}{2}R^{\gamma}_{m\delta
l}\nabla^{m}\nabla^{l}f_{G}\right]\epsilon_{\epsilon\gamma\chi}\epsilon^{\epsilon\delta}_{\psi}\\\nonumber&-&4R^{lm}\nabla_{m}\nabla_{l}h_{\psi\chi}f_{G}
+2Rh_{\psi\chi}\Box f_{G}
+\frac{1}{3}\left[\left(\varrho+p\right)f_{T}+4R^{m\alpha}R_{m\alpha}f_{G}\right.\\\nonumber&+&\left.4R^{lm}R^{\alpha}_{l\alpha
m}f_{G}-2R^{2}f_{G}-2R^{\beta}_{lmn}R^{lmn}_{\beta}f_{G}-2R\Box
f_{G}\right.\\\nonumber&+&\left.16R^{lm}\nabla_{m}\nabla_{l}f_{G}-4R^{m\alpha}\nabla_{\alpha}\nabla_{m}f_{G}
-4R^{m\beta}\nabla_{\beta}\nabla_{m}f_{G}\right.\\\nonumber&-&\left.4R^{\alpha}_{l\alpha
m}\nabla^{m}\nabla^{l}f_{G}\right]h_{\psi\chi}+\frac{1}{6}fh_{\psi\chi},\\\nonumber
\textsf{O}^{GT} &=&\left[\frac{1}{2}R_{m\epsilon}R^{m
p}f_{G}+\frac{1}{2}R^{lm}R^{p}_{m\epsilon
l}f_{G}-\frac{1}{4}RR^{p}_{\epsilon}f_{G}-\frac{1}{4}R_{\epsilon
lmn}R^{lmn p}f_{G}\right.\\\nonumber&+&\left.\frac{1}{2}
R^{p}_{\epsilon}\Box
f_{G}+\frac{1}{4}R\nabla^{p}\nabla_{\epsilon}f_{G}-\frac{1}{4}R^{lp}\nabla_{\epsilon}\nabla_{m}f_{G}-\frac{1}{2}R^{m}_{\epsilon}\nabla^{p}\nabla_{m}
f_{G}\right.\\\nonumber&-&\left.\frac{1}{2}R^{p}_{m\epsilon
l}\nabla^{m}\nabla^{l}f_{G}\right]g^{\psi\chi}\epsilon_{p
\delta\chi}\epsilon^{\epsilon\delta}_{\psi}+\left[-\frac{1}{2}R_{m\delta}R^{lp}f_{G}-\frac{1}{2}R^{lm}R^{p}_{m\delta
l}f_{G}\right.\\\nonumber&+&\left.\frac{1}{4}RR^{p}_{\delta}f_{G}+\frac{1}{4}R_{\delta
lmn}R^{lmnp}f_{G}-\frac{1}{2}R^{p}_{\delta}\Box
f_{G}-\frac{1}{4}R\nabla^{p}\nabla_{\delta}f_{G}\right.\\\nonumber&+&\left.\frac{1}{4}R^{lp}\nabla_{\delta}\nabla_{m}f_{G}+\frac{1}{2}
R^{m}_{\delta}\nabla^{p}\nabla_{m}f_{G} +\frac{1}{2}R^{p}_{m\delta
l}\nabla^{m}\nabla^{l}f_{G}\right]g^{\psi\chi}\epsilon_{p\epsilon
\chi}\epsilon^{\epsilon\delta}_{\psi}\\\nonumber&+&\left[-\frac{1}{2}R_{m\epsilon}R^{m\gamma}f_{G}-\frac{1}{2}R^{lm}R^{\gamma}_{m\epsilon
l}f_{G}
+\frac{1}{4}RR^{\gamma}_{\epsilon}f_{G}+\frac{1}{4}R_{\epsilon
lmn}R^{lmn\gamma}f_{G}\right.\\\nonumber&-&\left.\frac{1}{2}R^{\gamma}_{\epsilon}\Box
f_{G}-\frac{1}{4}R\nabla^{\gamma}\nabla_{\epsilon}f_{G}+\frac{1}{4}R^{m\gamma}\nabla_{\epsilon}\nabla_{m}f_{G}
+\frac{1}{2}R^{m}_{\epsilon}\nabla^{\gamma}\nabla_{m}
f_{G}\right.\\\nonumber&+&\left.\frac{1}{2}R^{\gamma}_{m\epsilon
l}\nabla^{m}
\nabla^{l}f_{G}\right]g^{\psi\chi}\epsilon_{\delta\gamma\chi}\epsilon^{\epsilon\delta}_{\psi}+\left[\frac{1}{2}R_{m\delta}R^{m\gamma}f_{G}
+\frac{1}{2}R^{lm}R^{\gamma}_{m\delta
l}f_{G}\right.\\\nonumber&-&\left.\frac{1}{4}RR^{\gamma}_{\delta}f_{G}-\frac{1}{4}R_{\delta
lmn}R^{lmn\gamma}f_{G}+\frac{1}{2}R^{\gamma}_{\delta}\Box
f_{G}+\frac{1}{4}R\nabla^{\gamma}\nabla_{\delta}f_{G}\right.\\\nonumber&-&\left.\frac{1}{4}R^{m\gamma}\nabla_{\delta}\nabla_{m}f_{G}
-\frac{1}{2}R^{m}_{\delta}\nabla^{\gamma}\nabla_{m}f_{G}-\frac{1}{2}R^{\gamma}_{m\delta
l}\nabla^{m}\nabla^{l}f_{G}\right]g^{\psi\chi}\epsilon_{\epsilon\gamma\chi}\epsilon^{\epsilon\delta}_{\psi}\\\nonumber&-&
12R^{lm}\nabla_{m}\nabla_{l}h_{\psi\chi}f_{G} +6R\Box f_{G}
+\left[\left(\varrho+p\right)f_{T}+4R^{l\psi}R_{l\psi}f_{G}\right.\\\nonumber&+&\left.4R^{lm}R^{\psi}_{l\psi
m}f_{G}-2R^{\chi}_{lmn}R^{lmn}_{\chi}f_{G}-2R^{2}f_{G}-2R\Box
f_{G}\right.\\\nonumber&+&\left.16R^{lm}\nabla_{l}\nabla_{m}f_{G}
-4R^{l\chi}\nabla_{\chi}\nabla_{l}f_{G}-4R^{l\psi}\nabla_{\psi}\nabla_{l}f_{G}\right.\\\nonumber&-&\left.4R^{\psi}_{l\psi
m}\nabla^{l}\nabla^{m}f_{G}\right]+\frac{1}{2}f.
\end{eqnarray}\\
{\bf Data Availability Statement:} This manuscript has no associated
data.

\end{document}